\documentclass[prl,twocolumn,showpacs]{revtex4}

\usepackage{graphicx}
\usepackage{hyperref}
\usepackage{amssymb,amsmath}
\usepackage{epstopdf}

\begin{document}

\title{Squeezing at a telecom wavelength, a compact and fully guided-wave approach}

\author{F. Kaiser$^1$, B. Fedrici$^1$, A. Zavatta$^{2,3}$, V. D'Auria$^1$}
\email{virginia.dauria@unice.fr}
\author{S. Tanzilli$^1$}
\affiliation{$^1$Universit\'e Nice Sophia Antipolis, Laboratoire de Physique de la Mati\`ere Condens\'ee, CNRS UMR 7336,  Parc Valrose, 06108 Nice Cedex 2, France\\ 
$^2$Istituto Nazionale di Ottica (INO-CNR) Largo Enrico Fermi 6, 50125 Firenze, Italy\\
$^3$LENS and Department of Physics, Universit\'a di Firenze, 50019 Sesto Fiorentino, Firenze, Italy}

\begin{abstract}
We demonstrate, for the first time, the realization of an entirely guided-wave squeezing experiment at a telecom wavelength. The state generation relies on waveguide non-linear optics technology while squeezing collection and transmission are implemented by using only telecom fibre components. We observe up to $-1.83\pm0.05$\,dB of squeezing emitted at 1542\,nm in CW pumping regime. The compactness and stability of the experiment, compared to free-space configurations, represent a significant step towards achieving out-of-the-lab CV quantum communication, fully compatible with existing telecom fibre networks. We believe that this work stands as a promising approach for real applications as well as for "do-it-yourself" experiments.
\end{abstract}

\date{\today}
\maketitle

Generation and manipulation of continuous variable (CV) non-classical states of light are the object of intense research due to their importance in both fundamental and applied physics \cite{Laurat2006, Zavatta2010}. Among others, a valuable feature of CV quantum resources is that they can be generated in a deterministic way at the output of non-linear optical media \cite{Leuchs2010}. Moreover, CV entanglement is affected but never vanishes completely for any level of external loss \cite{Braunstein2005}. On these bases, CV quantum optics has experienced an increasing interest for its application to quantum key distribution (QKD) \cite{CerfBook2010}, with many proposals based on both single-mode \cite{Gottsman2001, Filip2011, Werner2013} and two-mode squeezed light \cite{Andersen2012, Diamanti2013}. Entanglement distillation and entanglement swapping schemes for long distance quantum communication have been demonstrated \cite{CerfBook2010, Sasaki2010} and systematic studies have been performed on the robustness of non-classicality against the communication channel losses \cite{Solimeno2012, Ziman2014}. \\
A further step towards real-world applications of CV quantum communication has been done by generating squeezed light in the telecom C-band of wavelengths, where low-loss optical fibres and high performance standard components are available \cite{Schnabel2011, Peng2011, Peng2013}. In the perspective of implementing quantum networks that exploit optical fibres to connect distant atomic quantum memories, a quantum interface has recently been developed, converting squeezed light from telecom to visible wavelengths compatible with suitable atomic transitions \cite{Schnabel2015}. In the same spirit, a light-matter interface, coupling light guided in a tapered nanofibre to cold atoms, has been demonstrated \cite{Laurat2015}. \\

In this framework, and in order to comply with further out-of-the-lab realizations of CV quantum optics, we demonstrate, for the first time, the feasibility of a full guided-wave approach for both the generation and measurement of squeezed light at a telecom wavelength. \\
In our scheme, single-mode squeezing at 1542\,nm is generated by spontaneous parametric down conversion (SPDC) in a periodically poled lithium niobate ridge-waveguide (PPLN/RW). At the output of the PPLN/RW, the non-classical beam is measured with a fibre homodyne detector. This configuration allows implementing an extremely easy setup, entirely based on commercially available components, and fully compatible with existing fibre networks. On one hand, non-linear optics based on waveguide technology offers, compared to bulk implementations, better compactness and stability \cite{ReviewSeb}, as well as the possibility of efficient SPDC in a single pass arrangement \cite{Fejer2002, Furusawa2007, Hirano2008}. On the other hand, the use of off-the-shelf telecom fibre components permits the realization of a simple and plug-and-play setup that requires no alignment effort for spatial mode matching and that can be straightforwardly modified by connecting additional fibre components \cite{Lutfi2015}. Our approach, by first combining these two technologies and CV quantum optics, paves the way towards the realization of accessible and versatile experiments for CV quantum communication \cite{Pirandola2012}. \\
Eventually, we note that the possibility of miniaturizing CV quantum technologies is at the center of recent investigations, in particular with the demonstration of a photonic chip for \emph{in situ} operations on squeezing \cite{Furusawa2015}. Compared to this work, our paper follows a complementary vision by fully exploiting guided-wave optics for networking between remote quantum sources and nodes.
\\
\begin{figure*}[htbp]
\centering
\fbox{\includegraphics*[width=2\columnwidth]{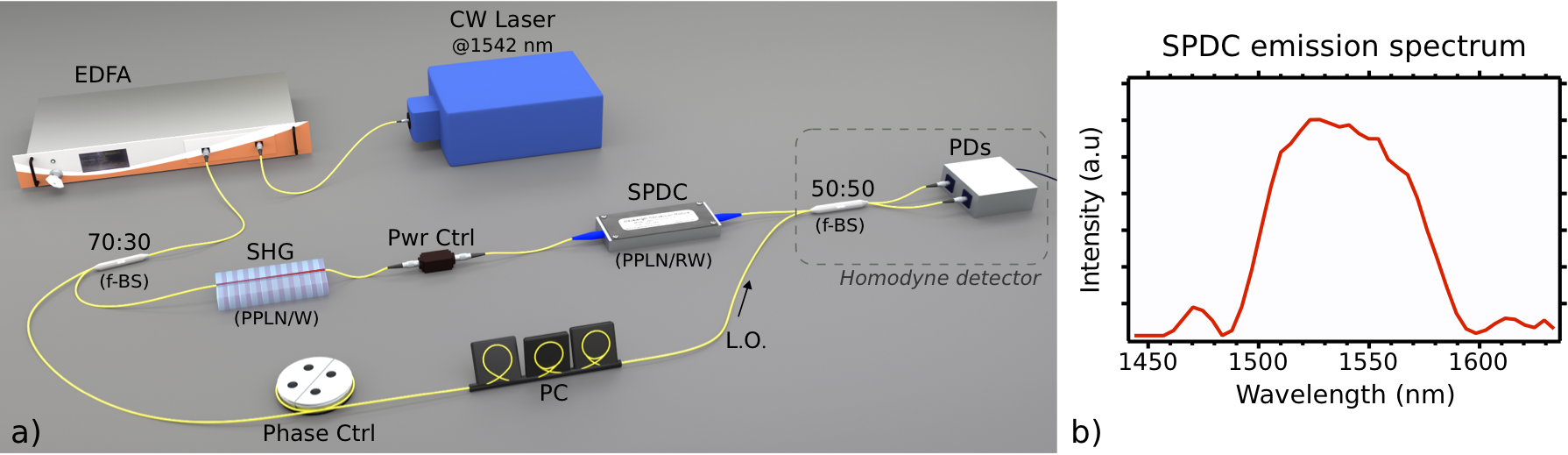}}
\caption{a) Schematic of the experimental setup. A fiber coupled CW telecom laser at 1542\,nm is amplified (EDFA) and split into two beams by means of to a 70:30 fibre beam splitter. The less intense beam serves as the local oscillator (LO) while the brighter one is frequency doubled via SHG in a PPLN/W and used to pump a ridge waveguide (SPDC, PPLN/RW). The power of the beam at 771\,nm is controlled with an in-line fibred attenuator (Pwr Ctrl). At the output of the SPDC stage, the squeezed vacuum state at 1542\,nm is sent towards a fibre homodyne detector where it is optically mixed with the LO based on a balanced fibre beam splitter (50:50) followed by InGaAs photodiodes (PDs). The LO phase is scanned thanks to a fibre-stretcher module (Phase Ctrl) while a fibre polarization controller (PC) allows the polarization mode matching at the homodyne detector. b) SPDC emission spectrum. The spectrum is centered at 1542\,nm and shows a bandwidth of 80\,nm FWHM, corresponding to 10\,THz.}
\label{setup}
\end{figure*}

The experimental setup is presented in Fig.\,\ref{setup}-a. A telecom continuous wave (CW) laser at 1542\,nm (\emph{Toptica}, DL Pro, fibre coupled) is amplified using an erbium doped fibre amplifier (EDFA, Keopsys, CEFA-C-HG) and directed towards a 70:30 fibre beam splitter. The less intense beam is used as the local oscillator (LO) for the homodyne detector. The brighter beam is frequency doubled to 771\,nm via second harmonic generation (SHG) to be subsequently employed as pump field for the SPDC process for the squeezing generation. The SHG stage is implemented employing a commercial periodically poled lithium niobate waveguide (PPLN/W, \emph{HC-photonics}, efficiency of $\sim 2000\%/W$). At its output, frequency doubled light is directly collected using a single mode fibre. The beam at 771\,nm then passes thought an in-line variable fibre attenuator and is directed towards the SPDC stage. Type-0 SPDC is obtained in a 4 cm-long commercial PPLN/RW working at frequency degeneracy around 1542\,nm (\emph{NEL}, WH-0770-000-F-B-C). We stress that this is, to our knowledge, the first time a PPLN/RW is employed for a CV experiment. The ridge waveguide structure offers a strong light confinement owing to a step index profile and guarantees high conversion efficiency over a large operation bandwidth of operation \cite{NTTridge}. By means of single-photon regime characterization, we estimate an SPDC conversion efficiency of $\sim 1.2\cdot10^6$ photon pairs/mW/GHz/s. The SPDC emission covers a continuous spectral bandwidth of 80\,nm FWHM (see Fig.\,\ref{setup}-b), corresponding to 10\,THz in the frequency domain. We stress that, as no optical cavity is required to enhance the SPDC process, this value represents directly the squeezing bandwidth \cite{Furusawa2007}. The PPLN/RW input and output facets are connected to polarization maintaining fibres (PMFs), whose coupling with the ridge is optimized by the manufacturer thanks to micro-lenses. For the input pump beam at 771\,nm, we measured a fibre-to-waveguide transmission of $\sim 0.43$. At the output of the ridge waveguide, single-mode squeezed light at 1542 nm is collected with a measured coupling efficiency $\eta_c \approx 0.80$ and it is directly available at the output of the PMF. To detect it, we send it to a fibre homodyne detector based on a 50:50 fibre beam splitter (50:50 f-BS) followed by two InGaAs photodiodes. In order to minimize Fresnel reflection losses, the outputs of the 50:50 f-BS are spliced to AR-coated fibre optic patchcords. The measured transmission of the signal path from the PMF output to the homodyne optical outputs is $\eta_T \approx 0.95$.\\

In addition to the compactness and stability of the setup, a major advantage of guided-wave optics lies in the achievement of a high degree of spatial mode matching between the LO and the signal without optical adjustment at the 50:50 f-BS inputs \cite{Furusawa2015}. This huge benefit extremely simplifies the homodyne detector implementation (see Fig. \ref{setup}-a). Moreover, compared to the pulsed regime, CW pumping bypasses the difficulties of obtaining a LO matched with squeezed light in the temporal domain \cite{Hirano2008}. Polarization matching is simply obtained by inserting a fibre polarization controller on the LO path. A home-made fibre stretcher implements the LO phase scanning \cite{Tanzilli2014}. At the 50:50 f-BS outputs, light is directly sent to two InGaAs photodiodes (\emph{Thorlabs}, FGA10, without cap) showing each $\eta_d \approx 0.88$ quantum efficiency at 1542\,nm. The difference photocurrent, obtained by connecting the photodiodes to each other, is amplified by a low-noise home-made transimpedence amplifier with bandwidth of $\sim$5\,MHz. The noise power is measured for each quadrature phase by an electronic spectrum analyzer (ESA, \emph{HP}, ESA-L1500A) set at zero-span around the analyzing frequency. As discussed in \cite{Furusawa2007}, the observed bandwidth of squeezed states produced by PPLN/Ws is ultimately limited by that of the homodyne detection. 
\begin{figure}[htbp]
\centering
\fbox{\includegraphics[width=0.95\linewidth]{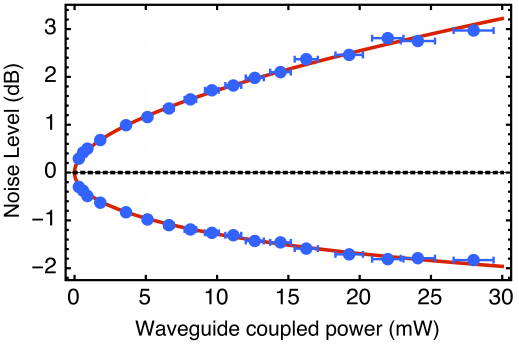}}
\caption{Squeezing and antisqueezing levels as functions of the pump power coupled inside the PPLN/RW. The error on each point is $\pm0.05$ dB for the noise levels and $\sim5\%$ on the pump powers. The data fit gives an overall detection efficiency of $\eta_{fit}=0.54\pm0.01$ and a squeezing parameter $\mu_{fit}=(0.101\pm0.002)\cdot\,mW^{-1/2}$ (see eq. \ref{Quad}).}
\label{sqvspower}
\end{figure}
In our experiment, we chose to work at 2\,MHz, where for a LO of 6.3\,mW, we observe a signal-to-noise ratio (SNR) of 15.6\,dB. Residual electronic noise associated with the detectors affects the measurement as an additional loss and can be taken into account by a factor $\eta_{el}=(SNR-1)/SNR \approx 0.97$ \cite{Lvovsky2007}. Note that squeezing at the telecommunication wavelength of 1550\,nm can be detected over bandwidths of more than 2\,GHz, by means of ultra-fast detectors and electronics \cite{Schnabel2013}. In this context, we recall that the exploitation of broadband squeezing is a key element for quantum channel multiplexing and high-speed quantum communication \cite{Schnabel2013, Treps2013}. \\

In the presence of losses, the variance of measured squeezed state quadratures, $\Delta X^{2}_m(\theta)$, can be written as \cite{Kuwamoto2005}: 
\begin{equation}
\Delta X^{2}_m(\theta)=\eta \left[e^{2\cdot r} cos(\theta)+e^{-2\cdot r} sin(\theta)\right]+1-\eta,
\label{Quad}
\end{equation}
where $\eta$ is the overall detection efficiency. Here, the values $\theta$=0 and $\theta$=$\pi$/2 correspond to anti-squeezing and squeezing, respectively. In equation (\ref{Quad}), the squeezing parameter $r$ depends on the SPDC pump power as $r=\mu*\sqrt P$, where $\mu$ is proportional to the crystal length and to the non-linear interaction strength.\\
Fig.\,\ref{sqvspower} shows squeezing and anti-squeezing (both in dB) as functions of the pump power at 771\,nm. Each point corresponds to an average over several acquisitions with an error of $\pm 0.05$\,dB. Data include the effect of imperfect ridge waveguide-to-fibre coupling, propagation losses in the waveguide and in the fibre components, non-unitary detection efficiency, and residual electronics noise. Pump powers reported on the graph are estimated with an error of $\sim5\%$ and refer to inferred values at the PPLN/RW input, taking into account the coupling and propagation losses at 771\,nm. Different power levels are obtained with the variable attenuator at the SHG output. As it can be seen, experimental data for both squeezing and anti-squeezing correctly follow the quadratic behavior predicted by the theory. This shows, in particular, the absence of unwanted excess noise on anti-squeezed quadratures. By fitting the entire ensemble of data with equation (\ref{Quad}), we obtain $\mu_{fit}=(0.101\pm0.002)\cdot$\,mW$^{-1/2}$ and an overall detection efficiency $\eta_{fit}=0.54\pm0.01$. A comparison of $\eta_{fit}$ with our estimated detection efficiency, $\eta_{est}=\eta_c \cdot \eta_T \cdot \eta_d  \cdot \eta_{el} \approx 0.65$, gives for the propagation losses inside the ridge waveguide $\eta_{wg}\simeq$0.4\,dB/cm, in agreement with typical reported values \cite{NTTridge}. 
\begin{figure}[htbp]
\centering
\fbox{\includegraphics[width=0.9\linewidth]{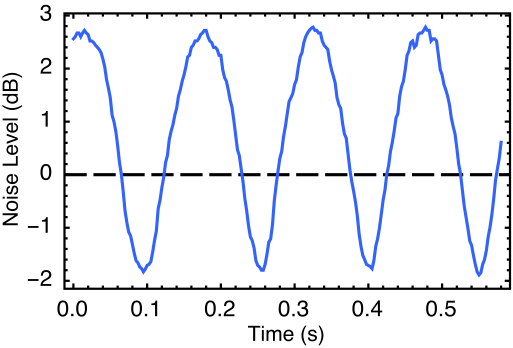}}
\caption{Normalized noise variances at 2\,MHz of the squeezed vacuum state at 28\,mW of coupled pump power as a function of the local oscillator phase (proportional to the time). The spectrum analyzer resolution and the video bandwidths are set to 300\,kHz to 30\,Hz, respectively.}
\label{squeezing}
\end{figure}
Fig.\,\ref{squeezing} shows a typical squeezing curve obtained by scanning the phase of the local oscillator overtime. It corresponds to the highest pump power (28\,mW) coupled inside the PPLN/RW and to a measured squeezing value of $-1.83\pm0.05$\,dB, with an anti squeezing of $2.79\pm0.05$\,dB. We stress that, by correcting the measured values for $\eta_{est}$, we can infer the squeezing at the output of the waveguide to be $\sim$-3.3\,dB, which is among the best values reported to date for CW-pumped squeezing \cite{Furusawa2007}. A further improvement could be obtained by reducing the propagation losses inside the PPLN/RW and by employing detectors with a higher $\eta_d$ at $\sim$1550\,nm. The fabrication of low loss ridge waveguides is at the center of intense investigations with encouraging results leading to propagation losses lower than $0.2$\,dB/cm \cite{RidgeLoss}. In parallel, detection efficiencies as high as $0.99$ at 1550\,nm have already been demonstrated on custom detectors \cite{Schnabel2011}. Eventually, the setup performances could be enhanced by improved pumping conditions. These could be achieved, for example, by implementing another PPLN/RW for the SHG-stage. This way, higher optical powers at 771\,nm could be achieved. Additionally, an improved fibre-to-waveguide coupling at the SPDC stage would lead to a greater efficiency for the squeezing generation.\\

Let us emphasize that for any application relying on entanglement as a quantum ressource, our setup can be easily modified so as to generate two-mode squeezing. This can be achieved by mixing the output of two identical PPLN/RWs at a 50:50 f-BS plugged just before the homodyne detector. The use of such a fibre component would automatically guarantee the mode-matching conditions for high-visibility quantum interference between the two squeezed states \cite{Furusawa2015} and will only introduce additional propagation loss typically of $0.05$\,dB. By considering our measured best squeezing level of $-1.83\pm0.05$\,dB and based on the Duan criterion for CV entanglement, this would lead to a correlation variance of $0.68<1$, which is well below the classical limit and in line with recent experiments showing the on-chip detection of Einstein-Podolsky-Rosen entanglement \cite{Furusawa2015}. \\

In conclusion, based on advanced waveguide non-linear optics and telecom technology, we have implemented for the first time an entirely guided-wave optical setup allowing both the generation and the detection of single-mode squeezed light at a telecom wavelength. We observe squeezing levels down to of $-1.83\pm0.05$\,dB. Our setup emploies plug-and-play components fully compatible with existing telecom fibre networks and requiring no alignment procedures for spatial mode matching. These advantages guarantee an extreme reliability and make our approach a valuable candidate for real-world quantum communication based on continuous variable quantum optics.

\section*{Acknowledgments}

 The authors acknowledge financial support from the Agence Nationale de la Recherche (ANR) through the $\emph{SPOCQ}$ (ANR-14-CE32-0019) and $\emph{CONNEQT}$ (ANR-2011-EMMA-0002) projects. A. Z. acknowledges financial support from the Universit\'e Nice Sophia Antipolis. F. K. acknolewdges financial support from the Foundation Simone et Cino Del Duca (Institut de France).The authors would like to thank Laurent Labont\'e, Philippe Bouyer, and  Baptiste Battelier for technical support.



\begin{thebibliography}{0}
\bibitem{Laurat2006} J. Laurat, G. Keller, J. A. Oliveira-Huguenin, C. Fabre, T. Coudreau, A. Serafini, G. Adesso, F. Illuminati, J. Opt. B \textbf{7}, S577  (2005).
\bibitem{Zavatta2010} M. Bellini and A. Zavatta, Progr. Optics  \textbf{55}, 41 (2010).
\bibitem{Leuchs2010}  U. L. Andersen, G. Leuchs, and C. Silberhorn, Laser \& Photon. Rev.  \textbf{4}, 337--354 (2010). 
 \bibitem{Braunstein2005} S. L. Braunstein, P. van Loock, Rev. Mod. Phys. \textbf{77}, 513 (2005).
\bibitem{CerfBook2010}  N. J. Cerf, G. Leuchs , E. S. Polzik, Quantum Information With Continuous Variables of Atoms and Light, Imperial College Press (2010).
\bibitem{Gottsman2001} D. Gottesman and J. Preskill, Phys. Rev. A \textbf{63}, 022309 (2001).
\bibitem{Filip2011} V. C. Usenko and R. Filip, New J. Phys. \textbf{13}, 113007  (2011).
\bibitem{Werner2013} T. Eberle, V. Handchen, J. Duhme, T. Franz, F. Furrer, R. Schnabel and R. F. Werner, New J. Phys. \textbf{15}, 053049 (2013).
\bibitem{Andersen2012} L. S. Madsen, V. C. Usenko, M. Lassen, R. Filip, and U. L. Andersen, Nat. Commun. \textbf{3}, 1083 (2012). 
\bibitem{Diamanti2013}  P. Jouguet, S. Kunz-Jacques, A. Leverrier, P. Grangier, and E. Diamanti, Nat. Photon. \textbf{7}, 378--381 (2013). 
\bibitem{Sasaki2010} H. Takahashi, J. S. Neergaard-Nielsen, M. Takeuchi, M. Takeoka, K. Hayasaka, A. Furusawa, and M. Sasaki, Nat. Photon. \textbf{4}, 178 --181 (2010).
\bibitem{Solimeno2012} D. Buono, G. Nocerino, A. Porzio, and S. Solimeno, Phys. Rev. A \textbf{86}, 042308 (2012).
\bibitem{Ziman2014} S. N. Filippov and M. Ziman, Phys. Rev. A \textbf{90}, 010301(R) (2014).
\bibitem{Schnabel2011} M. Mehmet, S. Ast, T. Eberle, S. Steinlechner, H. Vahlbruch, and R. Schnabel, Opt. Express \textbf{19}, 25763 (2011).
\bibitem{Peng2011} F. Y. Hou, L. Yu, X. J. Jiaa, Y. H. Zheng, C. D. Xie, and K. C. Peng, Eur. Phys. J. D \textbf{62}, 433--437 (2011).
\bibitem{Peng2013}  J. Zhao, X. M. Guo, X.Y. Wang, N. Wang, Y. M. Li, K. C. Peng, Chin. Phys. Lett. \textbf{30}, 060302 (2013).
\bibitem{Schnabel2015} C. Baune, J. Gniesmer, S. Kocsis, C. E. Vollmer, P. Zell, J. Fiurasek, and R. Schnabel, e-print arXiv: 1510.00603 (2015)
\bibitem{Laurat2015} B. Gouraud, D. Maxein, A. Nicolas, O. Morin, and J. Laurat, Phys. Rev. Lett. \textbf{114}, 180503 (2015).
\bibitem{ReviewSeb} S. Tanzilli, A. Martin, F. Kaiser, M. P. De Micheli, O. Alibart, and D. B. Ostrowsky, Laser \& Photon. Rev. \textbf{6}, 115 (2012).
\bibitem{Fejer2002} G. S. Kanter, P. Kumar, R. V. Roussev, J. Kurz, K. R. Parameswaran, and M. M. Fejer, Opt. Express \textbf{10}, 177 (2002).
\bibitem{Furusawa2007} K. Yoshino, T. Aoki, and A. Furusawa, Appl. Phys. Lett. \textbf{90}, 041111 (2007).
\bibitem{Hirano2008} Y. Eto, T. Tajima, Y. Zhang, and T. Hirano, Opt. Express \textbf{16}, 10650 (2008).
\bibitem{Lutfi2015} L. A. Ngah, O. Alibart, L. Labont\'e, V. D'Auria, S. Tanzilli, Laser \& Photon. Rev. \textbf{9}, L1 (2015).
\bibitem{Pirandola2012} C. Weedbrook, S. Pirandola, R. Garcia-Patron, N. J. Cerf, T. C. Ralph, J. H. Shapiro, and S. Lloyd, Rev. Mod. Phys. \textbf{84}, 621 (2012).
\bibitem{Furusawa2015} G. Masada, K. Miyata, A. Politi, T. Hashimoto, J. L. O'Brien and A. Furusawa, Nat. Photon. \textbf{9}, 316 (2015).
\bibitem{NTTridge} T. Umeki, O. Tadanaga, and M. Asobe, IEEE J. Quantum Electron. \textbf{46}, 1206 (2010).
\bibitem{Tanzilli2014} F. Kaiser, L. A. Ngah, A. Issautier, T. Delord, D. Aktas, V. D'Auria, M. P. De Micheli, A. Kastberg, L. Labont\'e, O. Alibart, A. Martin, S. Tanzilli, Opt. Comm. \textbf{327}, 7--16 (2014).
\bibitem{Lvovsky2007} J. Appel, D. Hoffman, E. Figueroa, and A. I. Lvovsky, Phys. Rev. A \textbf{75}, 035802 (2007).
\bibitem{Schnabel2013} S. Ast, A. Samblowski, M. Mehmet, S. Steinlechner, T. Eberle, and R. Schnabel,  Opt. Lett. \textbf{37}, 2367 (2012).
\bibitem{Treps2013} J. Roslund, R. M. de Araujo, S. Jiang, C. Fabre, and N. Treps, Nat. Photon. \textbf{8}, 109 (2013). 
\bibitem{Kuwamoto2005} T. Hirano, K. Kotani, T. Ishibashi, S. Okude, and T. Kuwamoto, Opt. Lett  \textbf{30}, 1722 (2005).
\bibitem{RidgeLoss} A. Gerthoffer, C. Guyot, W. Qiu, A. Ndao, M.-P. Bernal, N. Courjal, Opt. Mater. \textbf{38}, 37--41 (2014).

\end{thebibliography}
\end{document}